# Towards Modeling Human Attention from Eye Movements for Neural Source Code Summarization

AAKASH BANSAL, University of Notre Dame, USA
BONITA SHARIF, University of Nebraska - Lincoln, USA
COLLIN MCMILLAN, University of Notre Dame, USA

Neural source code summarization is the task of generating natural language descriptions of source code behavior using neural networks. A fundamental component of most neural models is an attention mechanism. The attention mechanism learns to connect features in source code to specific words to use when generating natural language descriptions. Humans also pay attention to some features in code more than others. This human attention reflects experience and high-level cognition well beyond the capability of any current neural model. In this paper, we use data from published eye-tracking experiments to create a model of this human attention. The model predicts which words in source code are the most important for code summarization. Next, we augment a baseline neural code summarization approach using our model of human attention. We observe an improvement in prediction performance of the augmented approach in line with other bio-inspired neural models.

## 1 INTRODUCTION

Source code summarization is the task of writing natural language descriptions of source code. These descriptions are called "summaries" and are a key component of software documentation for programmers. A programmer may read a short summary like "takes a screenshot" to quickly understand what a section of code does, without resorting to reading the source code. Despite the usefulness of these summaries, programmers often neglect to write or update them. The result is that automatic source code summarization has long been an appetizing target in software engineering research. The scientific community has long sought to enable machines to understand code in the way people do, so that those machines can describe code like a person would.

A confluence of recent advances in both software engineering and machine learning research is bearing fruit, such that automated code summarization seems almost within reach. In particular, *neural* source code summarization has held the vanguard of the state of the art since around 2017. Neural code summarization refers to approaches based on neural networks, namely the encoder-decoder architecture [61]. The encoder-decoder architecture is borrowed from problem domains such as image captioning, in which an encoder creates a vectorized representation of an image, and a decoder creates a representation of a natural language caption of that image. After training with sufficiently large datasets, an encoder-decoder architecture learns to identify features in the encoder that are associated with words in the decoder. Models for code summarization typically treat the code as the encoder input and the summary as the decoder. The idea is that the encoder learns features in the source code associated with words in the summaries.

The linchpin of almost all neural code summarization models is the "attention mechanism". An attention mechanism is the part of the model that emphasizes features in the encoder that are associated with features in the decoder. Metaphors in other problem domains are abundant: In an image captioning system, the attention mechanism will learn to connect the word "dog" to shapes in the image similar to a dog. In a machine translation system, the attention mechanism will learn to connect the English word "dog" to an e.g. German word "hund." In source code summarization, the attention mechanism connects words like "add" or "remove" to features in the code related to adding or removing data. The point is that the machine learns to pay attention to some features more than others in the encoder's representation.

Humans also pay attention to some features more than others. When human programmers read code, they tend to skim the code for select information they need to understand it, such as the return type and parameter list of a subroutine, or the variables that regulate access to a conditional block. One way to measure the attention that humans pay to source code is by tracking the movement of a person's eyes when reading that code. A plethora of eye tracking studies has demonstrated consistent processes that programmers follow when reading code [1, 2, 13, 47, 53], and it has even been demonstrated that programmers who are blind seek the same information from code, just using a different mechanical process [5]. In general, people are very efficient at extracting key features they need to understand code.

Models of human attention are a growing area in AI research due to the high-level intuition and experience that human attention reflects [42]. These models are occasionally called "biologically inspired" in that they attempt to mimic the processes observed in biological systems, namely the human eyes or brain. For example, [20] use fMRI scans to build a model of part of the human visual system to improve image classification. [27] model human attention to augment feature detection in images. These approaches attempt to model visual saliency to extract features most important to humans, when accomplishing the same tasks. The gains from these approaches tend to be modest (1-2% improvement is a typical expectation [20]), but are of high scientific value due to the new knowledge of human cognition and future potential of more human-like machines. This paper makes the following contributions.

- We create a model of human attention from eye movements during source code summarization. The model predicts where a person will look when reading source code.
- We use the model to generate predicted human attention for a different dataset of 190K Java methods.
- We use the predictions to augment the attention mechanism of a baseline neural code summarization approach.
- We evaluate the augmented model and show an improvement in line with other bio-inspired neural models.

In this paper, we attempt to model visual saliency, as observed by programmers when trying to summarize source code. We do this in way that is helpful for improving automatic source code summarization. We create a model of human attention during source code summarization, and we use that model to augment the attention mechanism of a baseline neural code summarization approach. We trained our model of human attention using eye tracking data published at ICSE [49].

To reduce experimental variables and maximize reproducibility, we augmented a "vanilla" neural encoder-decoder code summarization approach that is used as a baseline in many papers. We evaluate our model of human attention and the augmented code summarization technique. We observe a small but significant improvement by using human attention. We view this improvement as an important milestone in the design of neural models of code that behave more like people.

## 2 BACKGROUND & RELATED WORK

This section discusses key areas of background and related work, namely source code summarization, bio-inspired learning, and studies in software engineering (SE) literature using eye tracking.

### 2.1 Source Code Summarization

"Source code summarization" is a term coined by Haiduc *et al.* [23] around 2010 to refer to the task of writing short, natural language descriptions of source code [35, 61]. Code summarization is broadly related to the area of automatic documentation generation, in that the final destination of the summary descriptions is usually in documentation for programmers e.g., JavaDocs. The dream of writing code summaries automatically has floated among SE researchers for decades [21], driving strong advancements especially within the last ten years.

Around 2017, advancements in neural network research presaged a second generation of code summarization approaches. This second generation uses neural models to learn and extract relevant features from source code, and then connect those features to words to use in an output summary. Almost all involve an encoder-decoder model, derived from related work in machine translation and image captioning [39]. Essentially, the decoder is a language model which learns to predict words used in the output summary, while the encoder learns relevant features in source code. Typically an attention mechanism connects features from the encoder to words from the decoder. A majority of current research focuses on creating better encoder representations, such as using graph-based representations of code structure [30, 33, 62] or encoding contextual information provided by other code artifacts [9, 25].

### 2.2 Biologically-Inspired Learning

This paper differs from current code summarization literature in that we take a biologically-inspired approach. An AI model is "biologically inspired" if its design, training, or configuration is based on observations of the living world [11, 19]. Bio-inspired models form the cutting edge in several areas of AI research, for example by modifying the training process for object recognition to be more consistent with data from fMRI scans of humans performing the same recognition [10, 20, 41, 57]. Promising evidence exists comparing human programmer and machine attention [42], albeit using a survey methodology instead of eye-tracking data as in this paper.

Research into bio-inspired AI models tends to follow a similar pattern. Given an activity that the research wishes to automate (for e.g., image classification, text filtering, code summarization), the pattern is: 1) make observations of human or animal behavior when performing the activity the researcher desires to automate, 2) build a model of that behavior that predicts "typical" behavior for that activity, and 3) use the predictions of that model to augment or optimize an existing model of the desired activity. For example, [20] show humans images of objects in an fMRI device, and collect maps of brain activity when viewing different objects. Then, they create a function that predicts what brain areas are highlighted for the different objects. The feature weights used in classification are then balanced using predictions from this function.

### 2.3 Visual Saliency and Human Attention

This paper attempts to extract salient features from source code, similar to several visual saliency approaches in computer vision. There are several potential use cases for these salient features, such as locating objects in images [34, 36], biometric authentication [28], and image classification [20]. There have been several studies that try to model the human behavior of picking out these salient features using eye gaze data. For example, [44] present a classic top down approach aided by captions to predict salient objects in images. The study by [59] found that goal relevance, the most important information to complete the task, correlates with the eye scan-path. In this study we use gaze-time data to extract salient words in the source code that we then use to augment and improve a classic code summarization model. A point articulated by [29] and expanded on by others [12, 16] is that many of the failures of current neural models occur because those models do not accurately mimic human attention and learning. The hope of current research is to set the groundwork for further improvements.

### 2.4 Eye Tracking in Software Engineering (SE) Research

Eye tracking has a long history in software engineering research, starting in 1990 when [15] published the first study of eye movements by programmers. A survey by [53] chronicles up to 2015 and is expanded upon by [37]. An exemplar body of work is led by Sharif *et al.*, with advancements in both knowledge of how programmers read code [1, 2, 13, 17] and practical uses for this knowledge [52, 54–56]. Eye tracking data is widely recognized as representing the attention given to different parts of source code by programmers.

This paper makes use of an existing dataset of eye tracking behavior provided by [46, 49] published in 2014 and described in Section 2.5. This paper uses that dataset to build a model of human attention and then uses synthesized attention to improve a neural code summarization model. To the best of our knowledge, this is the first work we are aware of that uses bio-inspired neural models for code summarization based on eye gaze.

### 2.5 Overview of the Dataset

This section provides an overview of the dataset that was collected during the eye-tracking study by Rodeghero *et al.*, the data and other artifacts for which can be accessed online at [48]. They disclosed that the eye tracker they used was a commercial Tobii TX300. The device had a resolution of 1920x1080 and collected data at a sampling rate of 120Hz. The participants for their study were 10 professional programmers with an average programming experience of 13.3 years. Programmers were tasked with reading source code of a Java method, and subsequently writing a short summary for each method. The participants were allotted 1 hour for the study. However, the time allotment per task was not limited. In total, we have access to 130 data points from 9 programmers, where each data point is a list of eye-tracking fixations over a Java method by one of the programmers. Fixations are defined as locations viewed by the participant for more than 100 milliseconds. The words (areas of interest) at each fixation point as seen by the participants were also recorded. The authors then used these fixations to compute total eye gaze time, i.e. the time each participant spent looking at a particular word token in the Java method.

## 3 MODELING HUMAN ATTENTION

We model "human attention" based on the total eye gaze time as described in Section 2.5. Our intent is to be consistent with decades of studies of eye tracking in SE literature (see Section 2.4), in that amount of time a person spends reading a token indicates the importance of that token to the person's comprehension of the code. Each token is a word in the sequence that represents the source code. People skim tokens that are not important, and carefully read tokens that are [6, 58].

Our idea, in line with other bio-inspired models (see Section 2.2), is that a model that predicts eye gaze for each token in code will model the human attention for those tokens. The "sections of code" we target are subroutines, which are also the target of most code summarization research, as well as several eye tracking studies. When we refer to the "tokens" in a subroutine, we mean the list of tokens after lexicalization of that subroutine. These are the tokens that a programmer actually sees on their screen.

## 3.1 Inputs and Outputs

**Inputs - Abstract Syntax Tree** (AST): One challenge in modeling human attention to code is that humans read code as tokens (for sighted persons, these are the tokens visible to their eyes), while machine representations of code tend to rely on structures such as AST, an example of which is found in Figure 3. For example, a person may read an identifier name defined in the parameter list versus that identifier's use in a conditional statement. The person knows what the name means based on its location. A machine will represent that meaning as the name's position in the AST. For example, the identifier name may be related to a *parameter_list* node in the AST. That *parameter_list* node is not visible to the person. Practically, the machine and the person have the same information but in a different form. We follow three steps to convert source code into AST inputs for our model:

(1) We parse the source code through srcML [14] to generate AST nodes. These are nodes that surround the code tokens but are not seen by humans as described above.
(2) We perform a Depth-First-Search to traverse the AST, which gives us a sequence of nodes. This sequence is the "AST Nodes" input of prediction model described in the next subsection.
(3) We preserve the structure of AST as an adjacency matrix. This matrix is the "AST Edges" of the prediction model.

**Output - Percent Total Gaze Time** (ptgt): The attention we aim to predict is the percent of time spent reading a subroutine that is spent on each token. We call this the Percent Total Gaze Time (ptgt) of the token. Formally, each subroutine $s$ consists of tokens $\{t_1...t_n\}$. Each token has a gaze time $g$ that is the total amount of time a person spends reading it. The value $g$ includes regressions: if a person looks at $t_1$ for 20 milliseconds, then at $t_2$ for 30ms, then back to $t_1$ for 5ms, $g_1 = 25$ and $g_2 = 30$. The total gaze time for a subroutine is $\sum_{i=1}^{n} g_i$. The percent total gaze time of the $j$th token in subroutine $s$ would be $ptgt(t_j) = (g_j/\sum_{i=1}^{n} g_i)$. Note that we predict the percent of gaze time on each token not the absolute gaze time. This technique mimics attention on a neural code summarization model which is a value between 0 and 1 for each token indicating the relative important of that token to others in the code.

## 3.2 Prediction Model

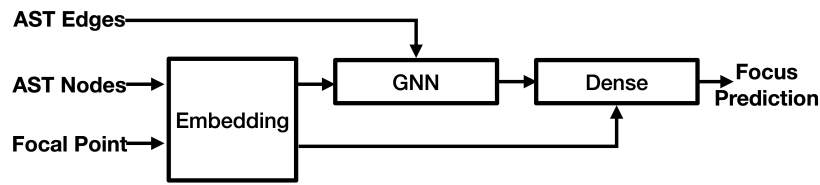

Fig. 1. Overview of the model predicting eye gaze time . The input includes the list of AST nodes, the edges among those nodes, and a focal point token in the AST. The output is a prediction of the gaze time a programmer would spend on that token.

Figure 1 shows an overview of our prediction model. The model is itself a neural network (not to be confused with the neural code summarization techniques in Section 5), and works as follows:

- First, we create a vector for each node in the AST sequence using an embedding layer.

- Meanwhile, also obtain the embedding vector of the focal point. We use the same embedding space as the AST nodes, because the focal point is itself a node in the AST. Finally, we concatenate the GNN output with the focal point embedding vector (results in an ($m$+1)x$n$ matrix), and feed them to a fully-connected output layer.
- Next, we use a graph neural network (GNN) to create a representation of the AST tokens. The initial states of the nodes for the GNN are the vectors from the embedding space. We then perform two iterations of the GNN, as recommended by [30] for GNN representations of source code. The output of the GNN is an $m$x$n$ matrix where $m$ is the number of nodes in the AST and $n$ is the embedding vector length (in this paper, $m$ is capped at 400 and $n$ is always 100, in line with recommendations of LeClair *et al.*).
- Finally, the output layer consists of just a single element, the *ptgt* of the focal point, which is a value between 0. We use Mean Squared Error (mse) as the loss function for our model. We derive the actual *ptgt* from the dataset of observed eye tracking values, which we explain in the next section.

We call the GNN model above **eye-gnn** throughout this paper. We also create an alternative we call **eye-rnn**. The eye-rnn model is identical to eye-gnn except that we replace the GNN with a recurrent neural network. We use a Gated Recurrent Unit (GRU) to model the AST nodes as a sequence. The inputs to the RNN model are AST tokens and the focal point, as it does not use the edges. Besides that, the model structure and output is the same as GNN. We present this alternative to provide a point of comparison to the GNN model.

Finally, we create a "pretrain" configuration for each model. In the standard configurations, the embedding space starts with random values (as is typical in neural models). In the pretrain configurations, we import the embedding space from a code summarization model published by [25]. We use the suffix -pretrain to indicate this configuration, e.g., **eye-gnn-pretrain** and **eye-rnn-pretrain**.

### 3.3 Data Preparation and Training

We trained our prediction model on data from the experiment detailed in Section 2.5. For each Java method, we used srcML [14] to extract the AST. Then, we created a sequence from the AST with the procedure mentioned in Section 3. We generated a vocabulary for this sequence using the vocabulary files published by [30] for AST sequences. We used this vocabulary so that it would be consistent with the vocabulary used in the baseline source code summarization technique we augment in Section 5. Next, we calculated the *ptgt* for each token that each programmer saw in the experiment. There is one *ptgt* value per token, per method, per programmer. Each of these values serves as one training example. Each token is the focal point, each method is the AST. Some methods are entered to the model more than once because some methods were viewed by multiple programmers.

We formed data examples from the focal point tokens only. We did not generate data examples for the structural AST nodes which the programmers could not have seen. Therefore, the prediction model does not learn to make predictions of the attention paid to structural AST nodes, since technically the programmers never saw those nodes. However, our model is still capable of learning to infer attention paid to those nodes, because those nodes are nearby the tokens that the people did see, as we point out in Section 3.2. Note that during the evaluation of our model of human attention, we held out some methods for a test set with a procedure detailed in Section 4. Also, when we evaluate our augmentation of the baseline code summarization approach (Section 6), we removed the six projects in this eye tracking dataset from the dataset used to train and test the baseline code summarization technique.

## 4 EVALUATING HUMAN ATTENTION

The objective of this experiment is to evaluate our prediction model for human attention. Recall that the input to our model is the AST of a subroutine and a focal point in that subroutine, and the output is the predicted *ptgt* for that focal point. We ask the following Research Questions (RQs) to evaluate our model:

> **RQ1** What is the best-performing configuration of our approach in terms of correlation of predicted attention to attention by human programmers?

The rationale behind RQ1 is that we have four configurations of our model (based on RNNs versus GNNs, and random start versus pretrained start), and these different configurations may have different performance. Performance differences may be attributed to the neural architecture (RNN versus GNN) because these architectures consider slightly different information. The RNN considers order of the AST nodes as a sequence only, while the GNN also considers AST edges. GNNs and RNNs have been shown to produce different representations of code [4, 30, 31], and it is possible that these representations will have a different impact on the prediction of human attention. Likewise, pretraining the word embedding and RNN layers may have an impact on predicted human attention.

### 4.1 Methodology

Our methodology for answering RQ1 is to calculate the Pearson correlation between the predictions from each configuration of our approach and each programmer in the eye tracking dataset, for each of the Java methods for which all programmers participated. We calculate correlation only over the visible tokens, not all nodes in the AST. Recall that the eye tracking dataset had nine programmers and 67 Java methods. We have eye tracking data for all nine programmers for four of these methods (we have less than nine programmers for the other methods). We trained each configuration of our approach using a "hold out one method and programmer" technique: Consider the set of nine programmers $P$, the set of four methods $T$, and remaining java methods $M$. We held out $p_i$ and $t_j$ as a test set, so the model saw eye tracking data for any method for $p_i$, and no eye tracking data for any programmer for $t_j$. Then, we trained on $P - p_i$ and $M + (T - t_j)$. That is, the model saw all methods except for $t_j$ and all programmers except for $p_i$. The result was 36 (4x9) folds: one for each $T$, for each $P$.

Next, we computed the predicted human attention for the method held out in each fold. We calculated the Pearson correlation between these predictions and the *ptgt* for each visible token in each held out method and programmer. The purpose of using correlation is that we aim to determine if the predicted attention paid to each focal point rises when the human attention rises. Note again that we do not aim to predict the actual gaze time in milliseconds – we aim to predict *ptgt*, which is a measure of the relative importance of each focal point. Positive correlation implies that the focal points that the model predicts are important, are actually viewed more by a person.

### 4.2 Results and Discussion

We found that `eye-gnn` produces the highest average Pearson correlation of predicted eye attention to actual eye attention. Figure 2 (a) shows this result over 100 training epochs. While all models achieve positive correlation that rises with the first few training epochs, `eye-gnn` consistently reaches the highest correlation, and also reaches the highest peak correlation of around 0.35. This result is as expected for this proof-of-concept study in predicting human attention to different areas of source code, even if 0.35 is generally considered only moderate Pearson correlation.

We also found that these results are generally consistent across different programmers and Java methods. Figure 2 shows the correlation of each of the nine programmers (labeled 1-10, skipping

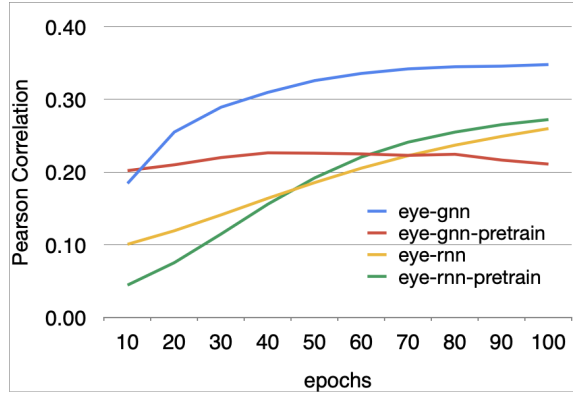

(a)

| | | \multicolumn{9}{c}{programmers} | |
|---|---|---|---|---|---|---|---|---|---|---|---|
| | | 1 | 2 | 3 | 4 | 5 | 7 | 8 | 9 | 10 | avg. |
| methods | A | N/A | .492 | .254 | .410 | .286 | .319 | .332 | .402 | .370 | .358 |
| | B | .254 | .43 | .053 | .553 | .383 | .375 | .367 | .182 | .484 | .342 |
| | C | −.12 | .435 | −.04 | −.04 | .443 | .255 | .402 | .337 | .123 | .199 |
| | D | .452 | .262 | .338 | .583 | .545 | .537 | .560 | .556 | .400 | .470 |
| m. avg. | | .195 | .405 | .151 | .376 | .414 | .371 | .415 | .369 | .344 | .338 |
| p. avg. | | .947 | .971 | .946 | .972 | .969 | .976 | .979 | .964 | .974 | .966 |

(b)

Fig. 2. (a) Average Pearson correlation of nine programmers for four configurations of the model. Note programmer 6 is missing because they were excluded by the original study [49]. The y-axis is average Pearson correlation. The x-axis denotes the epochs. (b) Pearson correlation scores between our attention prediction model (GNN, epochs 70) and actual programmer attention. The row *m. avg.* is the average Pearson correlation of each programmer to all four java methods we hold out to test. The row *p. avg.* is the Pearson correlation of each programmer to the average of all programmers.

6, to be consistent with the original dataset [49]) to predictions by eye-gnn for each of the four methods in the original paper's test set. For method C we observe negative correlation with three programmers. Method C could be a specially challenging method for the model to predict over. We observe low correlation overall with programmer 3 over all four test methods. This could be due to different code comprehension strategies [13, 47] and attention strategies [45] of programmers. A study by [5] found that there are some common elements that all programmers look for, even blind programmers. These are the salient features our model attempts to learn. Therefore, average correlations are more meaningful than individual correlations.

Note that we do not claim that eye-gnn is yet equivalent to a human programmer. Figure 2(b) also shows (under the label *p.avg.*) the correlation of programmers to each other. In general, the programmers' attention is very closely correlated to each other, with a Pearson correlation consistently above 0.94. The predictions from eye-gnn are less correlated at around 0.35. Given our small dataset we do not expect to perfectly predict human attention.

A likely explanation for the model's ability to learn to predict human attention lies in the AST. Consider the example source code in Figure 3 (which is method D in the test set). The training set is relatively small and the vocabulary is limited, meaning that the ability of the model to generalize based on words alone is limited. For example, consider the if statement on line 11. The only actual human visual attention available are the tokens e.g. if, is model, field. The string of words labeled

Raw Source Code:

```
         public static String[] getColumnNamesWithPrefix(Field field, String prefix) {
line 2 ->    Column c = field.getAnnotation(Column.class);

             if(c != null && c.value().length > 0) {
                 String[] cols = c.value();
                 for(int i=0;i<cols.length;i++)
                     cols[i]=prefix+cols[i];
                 return cols;
             }
line 11 ->   if(isModel(field.getType())) {
                 ClassInfo ci = getClassInfo(field.getType());
                 List<String> keys = new ArrayList<String>();
                 if(ci.keys.size()==1)
                     return new String[] { field.getName() };

                 for (Field key : ci.keys) {
                     keys.addAll(Arrays.asList(getColumnNamesWithPrefix(key,
                         prefix+field.getName()+"_")));
                 }

                 return keys.toArray(new String[keys.size()]);
             }

             return new String[]{ prefix + field.getName() };
         }
```

AST for line eleven:

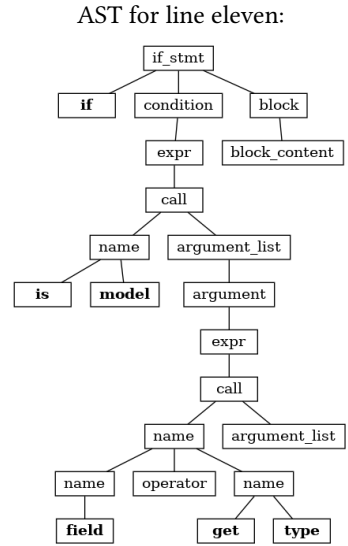

Actual and Predicted Attention on AST nodes for line 2:

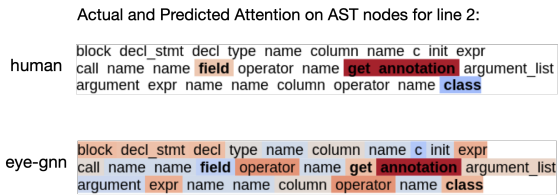

Actual and Predicted Attention on AST nodes for lines 11-12:

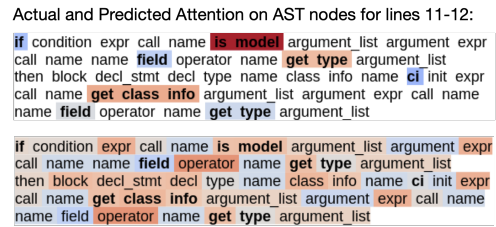

Fig. 3. Heatmaps comparing human and predicted attention for selected lines. We use gaze time normalized by total gaze time by programmer 5. We use red (hotter) to indicate higher attention while blue (cooler) indicates lower attention.

"human" for lines 11 and 12 correspond to the linearized AST for those lines. The colors indicate different amounts of attention (*ptgt*) by programmer 5 for the words on those lines. Actual human attention is zero to non-visible AST tokens because they exist only in the AST – the person never sees them. Yet the model is able to predict human attention for arbitrary methods that do not share vocabulary with methods in training set.

To understand how, consider the AST for line 11. When we train the model, the GNN propagates information from the visible tokens to the nearby AST tokens. What we observe is that the programmers are very likely to read the `is` and `model` tokens since they are the condition for an important `if` block. But the eye-gnn is also able to predict this situation because of the surrounding AST tokens. The model predicts higher attention to AST nodes like `argument_list`, which is propagated by the GNN to the `is` and `model` tokens. We observe similar behavior for `annotation` in line 2. The model works because it retains knowledge of the program's structure via the AST.

The AST also likely explains why the RNN and -pretrain models performed less well. The RNN propagates information via the order of tokens in the linearized AST, which is considered a less reliable way to model code than an AST [4]. Meanwhile, the pretrained models primarily benefit from increased information about the visible tokens, since the word embeddings are created from the visible tokens. These have only a limited effect because the vocabulary in the training set (of methods for which we have eye tracking data) is very small. The chief means by which the model learns seems to be through the AST. We also demonstrate the AST's effect in Section 6.4.

### 4.3 Threats to Validity

The key threats to validity of this experiment include: 1) the Java methods in the test group $T$, 2) the programmers in the test group $P$, and 3) the hardware and software inaccuracy inherent in all eye tracking datasets. We are bound by the available eye tracking data to limit the size of the test group. These methods and programmers were selected in the original experiment for a range of subject domain, experience level, and other factors to increase diversity, though a risk remains that the results of this experiment may change with different methods or programmers. Likewise, the hardware and software for eye tracking is not considered 100% accurate as the eye signal can get noisy at times. This is minimized by making sure calibration is done carefully and all best practices followed as indicated in the original experiment. While the original experiment [48] involved high-end research grade hardware and software, it is possible that our results could change due to these inaccuracies.

## 5 CODE SUMMARIZATION APPLICATION

This section shows how we augment an existing source code summarization approach with predictions of human attention. Our research objective is to determine the degree to which the human attention predictions impact the results of the baseline summarization model. We ask the following Research Question:

**RQ2** What is the performance of the code summarization approach, in terms of commonly-used evaluation metrics, when provided attention predictions from our human attention model configurations?

The rationale for RQ2 is that we created four different configurations of our human attention prediction model, and each may have a different impact on the baseline. We compare the best performing configuration eye-gnn to the baseline. Recall that the original eye tracking dataset from which we build our predictor of human attention was intended as a study of human programmers during code summarization. We showed in the previous section that our model's predictions correlate with actual human attention. That is an interesting finding academically, but it does not demonstrate how the predictions may be useful in practice. In this section, we show how predictions of human attention can be used to improve neural models of code summarization.

### 5.1 Augmenting an Existing Model

Our key augmentation to the baseline starts with predicting the human attention for each subroutine. Recall that we tested four configurations of our attention prediction model in the previous section (rnn, rnn-pretrain, gnn, gnn-pretrain). The output from each of these configurations is the predicted *ptgt* of a given focal point in a given subroutine. A key difference in the way we tested these configurations in the previous section and the way we use the models in this section is in the focal point. In the previous section, the focal points were the code tokens, since only the tokens are visible to the programmer. We could only compute correlation to these tokens. However, now we compute predicted attention for all elements in the sequence of AST nodes, including the tokens and the "invisible" AST structure nodes.

We compute attention to all AST nodes essentially because that is how machine attention works. The attention calculated in the machine's attention mechanism is computed over every element in the input sequence. Recall that the way the machine model represents meaning of a code token is largely via the neighboring structural nodes in the AST to the token ; i.e., a human knows what the token means by its position, but the machine learns this via the AST.

Note that there is a risk involved in predicting attention for all elements in the AST node sequence. The risk is that the attention prediction model will not have seen examples of these elements during training. The model will have to infer attention to these nodes without explicit training about them. This introduces a risk that the model will struggle to produce good predictions. However, in our view this risk is mitigated by the prediction model design. Recall that the attention model connects items in the sequence of AST nodes based on their connections in the AST (via a GNN). The result is that information from nodes representing code tokens (which are visible to programmers) will propagate to nearby AST nodes (which are not visible). The human attention prediction models have an opportunity to learn to predict attention to AST structural nodes due to their proximity to code tokens in the AST.

## 5.2 Model Design

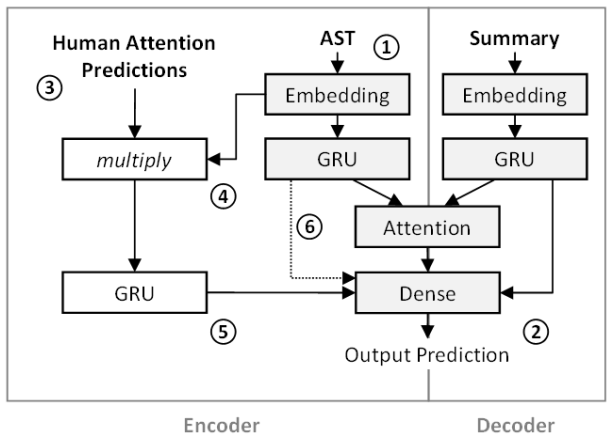

Fig. 4. Overview of how we augment an existing code summarization baseline. Gray components are the baseline. White components are our augmentations to the encoder.

Figure 4 shows an overview of how we augment the existing model. We start with a straightforward and popular baseline that is essentially just a vanilla encoder-decoder model. This baseline is marked as the gray components in the figure (areas 1 and 2). We provide the encoder the sequence of AST nodes from a subroutine (linearized from the AST as in Section 3.2). We provide the decoder the summary of the subroutine. We use the teacher forcing training procedure, like the vast majority of related work [31]. We use this vanilla encoder-decoder model because it is a baseline in many code summarization papers and because the model's relative simplicity reduces experimental variables.

The predictions from the human attention model from the last section serve as an additional input to the code summarization model (Figure 4, area 3). We use only one configuration at a time, so these predictions come from e.g. the gnn-pretrain configuration but not others. We normalize the predicted $ptgt$ values such that, for each subroutine, the mean $ptgt$ value is 1.0. Other $ptgt$ values will be above or below 1.0, indicating more or less "importance" of that word. Above average predicted attention for a word is indicated by a value greater than 1.0. The result is a "human attention vector" that contains a single value representing the predicted human attention for each node in the AST sequence.

We use this human attention vector to scale the embedding vectors in the sequence of AST nodes (Figure 4, area 4). The effect mimics how a machine attention mechanism typically functions: the embedding vectors of more important elements in the input sequence are emphasized, while the embedding vectors of less important elements are attenuated. The typical next step in a machine attention mechanism is to combine these vectors into a single vector representation. Finally, we concatenate the final state of this RNN to the input of the output layer of the baseline model. Our model augments representation using both a human attention and a machine attention mechanism.

## 5.3 Fair Baseline

We made one modification to the baseline to make it more "fair" for comparison. The modification is that, in the baseline only, we concatenated the output of the encoder GRU directly to the input of the output layer. The reason we made this modification is that otherwise the number of connections to the output layer would be much higher for the augmented model than for the baseline. It is possible that the augmented model would have improved prediction capacity simply because the network is larger: there are more "neurons" between the output dense layer and the previous parts of the model. This increased network size could be an unfair advantage to the augmented model. Therefore, to create as fair a comparison as possible, we increased the baseline model by an equal amount by concatenating the encoder's GRU final state to the output layer's inputs, as shown in Figure 4 area 6. The dashed line only exists in the baseline model.

## 6 EVALUATING CODE SUMMARIZATION

This section evaluates the code summarization approach from the previous section using the best performing configuration eye-gnn as input.

## 6.1 Methodology

We use a data-driven experimental methodology to answer RQ2. This methodology is identical to the procedure used in a vast majority of papers on neural source code summarization: we train different configurations of the neural model using a training set derived from a large repository of source code. We use the trained model to predict code summaries for subroutines in the test set derived from the repository. We compare the predictions to reference code summaries from the repository. We use three metrics for this comparison: METEOR [7], USE [51], and BLEU [43]. While BLEU has traditionally been the most popular metric, it has fallen under controversy in SE literature on code summarization: [50] show evidence strongly favoring METEOR over BLEU for metrics based on word overlap, while [51] show similar evidence favoring USE as a semantic similarity metric over BLEU. Therefore, we use METEOR and USE as primary metrics for evaluation, but still report BLEU to conform with past practice.

There is always a concern in neural models that performance differences could be due to random factors. We take two steps to help mitigate this concern. First, as recommended by [50], we use a paired t-test to test statistical significance of the difference of METEOR and USE, since these metrics have values suitable for measuring each output summary (unlike BLEU, which is only considered meaningful at corpus level). Second, we create a baseline of human attention vectors using random values. We create five random human attention vectors and use them as input in place of the output of our eye-gnn results. We report the minimum, maximum, and mean performance of these random vectors, and test the significance of our results against the best of the random vectors.

Note that we elected to conduct a data-driven experiment instead of a human study. Human studies rely on coarse-grained records of human perception (e.g., a 4-point Likert scale), which are noisy due to human factors such as bias and fatigue. These records have difficulty discerning small changes in performance such as the 1-2% range we might expect in a study of biologically-inspired neural models. The study by [50] found that improvements of up to 2 points (around 5-8% for our baselines) are not detectable by human evaluation. This does not mean that they are not important improvements, just that it would take a large human study to verify them. Such a study, with professional programmers that our research targets, would be cost-prohibitive. Instead, as they recommend, we conduct statistical significance tests to verify these improvements.

## 6.2 Dataset Augmented for Code Summarization

We create a dataset of 190k Java methods. We derived this dataset from a dataset of 2.1m Java methods published in [32], though we apply several filters recommended by [3] to improve quality and remove duplicates. Then we selected the top 10% of methods in terms of number of tokens per method. We selected this top 10% for two reasons. First, as [9] point out, a large portion of methods are very short (just 2-3 lines long) that often have trivial summaries (e.g., "plays music" for `playMusic()`). Second, the computational cost of inferring *ptgt* on every token in millions of methods is prohibitively high with current hardware. Focusing on the largest 10% of methods means we experiment with a challenging and interesting subset of methods in affordable time. We split the training, validation, and test sets from this dataset by project, in order to reduce the risk of information leakage from the training set into the test set [32]. We use a Java dataset because the original eye tracking data was performed using Java, and it is not clear that eye tracking data generalizes to another language.

## 6.3 Threats to Validity

The key threats to validity in this experiment include: 1) the dataset of Java methods, 2) the metrics we use for comparison, 3) the preprocessing and training procedures, and 4) the underlying human prediction models and training data for those models. The dataset is a threat to validity because different training, validation, and test examples could lead to different results and even different conclusions. We attempt to mitigate this threat by using a large dataset that has been vetted by multiple recent papers on neural code summarization. The metrics we use could be a threat because they guide our conclusions about performance of each model, and different metrics may rank models differently. We use METEOR and USE, since they are recommended by empirical data to help mitigate this threat, in preference to BLEU. We use srcML and teacher forcing as key parts of our preprocessing and training, but it is possible that different preprocessing scripts or training procedures would give different results. We use tools and procedures that are uncontroversial in the literature, yet it is still possible that different tools and procedures would give different results. Finally, the risks to the eye tracking study dataset still apply to this experiment.

## 6.4 Results and Discussion

Figure 5 summarizes our answer to RQ2. In short, the `eye-gnn` model of human attention leads to higher performance than the baseline. However, the degree of improvement is below a threshold where a human may immediately notice according to evidence [50].They point out that metric score improvements of less than 2 points may not be detectable in a user study. The problem is that a change not noticeable by humans could also be an artifact of measurement or due to random factors. Three pieces of evidence mitigate that risk:

First, we report improvements in three metrics instead of only one (most papers only report BLEU). As mentioned in Section 6.1, Roy *et al.* strongly favor METEOR to BLEU because METEOR is much more correlated with actual human programmer judgments. To provide more evidence, we also report improvements in USE, a metric found by [51] to also be favored over BLEU. All three metrics demonstrate a similar improvement.

Second, we follow the recommendation from Roy *et al.* to perform a paired t-test of the scores (where the predicted summary of each of two approaches are the pairs) for METEOR and USE that are meaningful at the summary level. BLEU is considered a corpus-level metric due to the brevity penalty, so a paired statistical test is not meaningful [18]. In any case, we found statistically-significant difference between `eye-gnn` and the baseline (Figure 5c).

Third, we found improvements over five configurations of our model in which we generated random values for the human attention vectors. In theory, it is possible that `eye-gnn` was merely

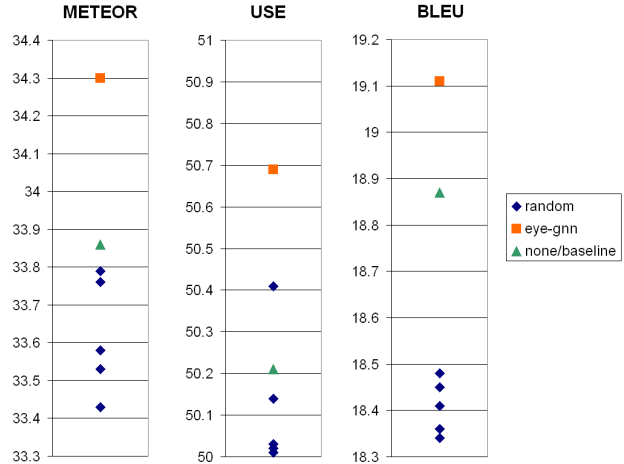

(a) Score comparison with different attention models.

|  | baseline | our models | random | | |
|---|---|---|---|---|---|
| human attn.: | none | eye-gnn | mean | max | min |
| BLEU | 18.87 | 19.11 | 18.41 | 18.48 | 18.34 |
| METEOR | 33.86 | 34.30 | 33.62 | 33.79 | 33.43 |
| USE | 50.21 | 50.69 | 50.12 | 50.41 | 50.01 |

(b) Comparison of peak scores for different attention.

|  |  | METEOR | | USE | |
|---|---|---|---|---|---|
|  |  | t stat | p-value | t stat | p-value |
| vs. baseline | eye-gnn | 2.52 | <0.01 | 2.71 | <0.01 |
|  |  |  |  |  |  |
| vs. random | eye-gnn | 3.08 | <0.01 | 3.27 | <0.01 |
|  | baseline | 0.43 | 0.33 | 0.35 | 0.36 |

(c) Paired t-test results for METEOR and USE.

Fig. 5. Key results of the code summarization experiment. First, (a) depicts model performance for different attention models, including with five random attention vectors. Then, (b) shows peak METEOR, USE, and BLEU scores for the code summarization baseline alone, the baseline when provided our human attention models, and the mean/max/min performance of the baseline when provided five random attention vectors. Last, (c) shows statistical significance tests.

a "lucky" input. Different studies have shown that randomly altering parameters in neural models can actually improve performance by inducing a regularization effect [22, 26, 60]. To help rule out this effect, we produced five random "human attention" vectors that have the same minimum and maximum values as eye-gnn. This approach of using randomly generated attention also used as a baseline against novel attention modelling approaches in other domains [44]. These five runs are visible as blue diamonds in Figure 5a. All are below eye-gnn. Note that a 1-2% improvement is closely in line with expectations from other code summarization research, so long as the improvement is clearly attributable to a single factor under controlled conditions, such as benefits from a particular representation of the code's structure [31]. This improvement is also in line with other bio-inspired [20, 38] and visual saliency prediction models [40, 44].

The examples in Figure 6 and 7 demonstrate how the benefits of predicted human attention may be borne out. In Figure 6 the method is from a Chess game program and creates a board configuration for test purposes. The name of the method indicates the use of the words "sets up the," but leaves open the question of *what* is set up. The baseline selects the word "pawn" from the text of the method, likely because it is a rare word that occurs several times. However, the approach using

Raw Source Code:
```
           protected void setUp() throws Exception {
               Knight knight = new Knight(5, 5, board, "White");
               board.add(knight);
               Pawn pawn1 = new Pawn(3, 4, board, "White");
               board.add(pawn1);
line 6 ->      Pawn pawn2 = new Pawn(4,5,board, "White");
               board.add(pawn2);
               Pawn pawn3 = new Pawn(7,5,board, "Black");
               board.add(pawn3);
               King king = new King(4,7,board, "Black");
               board.add(king);
               King king2 = new King(5,3,board, "White");
               board.add(king2);
               board.print();
               super.setUp();
           }
```

Example Method Summaries:

| | |
|---|---|
| *reference* | set up the board to test the upcoming methods |
| eye-gnn | set up the board |
| <none> | set up the pawn |

Predicted Human Attention on AST nodes for line 6:

eye-gnn    decl_stmt decl type name pawn name pawn2 init =
           expr operator new call name pawn argument_list (
           argument expr literal 4 , argument expr literal 5 ,
           argument expr name board ,
           argument expr literal " white " )

Fig. 6. Example method (ID 4479155 in reproducibility package). Reference, human-written summary at top, followed by the summaries generated by the model using eye-gnn and the baseline (without human attention prediction). The source code includes indicated line six, for which we visualize the predicted human attention by eye-gnn. Red and blue indicate different sides of the attention spectrum.

Example Method Summaries:

| | |
|---|---|
| *reference* | write the properties to the global properties |
| eye-gnn | write the properties to the current state |
| <none> | this method is called when the user selects the user |

Predicted Human Attention on AST nodes for line 2:

eye-gnn    unit function specifier public type name void
           name write properties parameter_list ( ) block
           { decl_stmt decl type specifier final name ide

Fig. 7. Example method (ID 22407475 in reproducibility package). The human-written, eye-gnn augmented and baseline summaries on the left. The predicted attention scores are on the right. Note, blue indicates lower than average attention and red indicates higher than average attention. The word "write" gets an above average attention score.

eye-gnn correctly uses the word "board." A possible reason emerges from the predicted human attention. The word "pawn" is activated to a notably different degree than "board." As we described in Section 5.1, this activation is multiplied with the word embedding vectors to raise or lower the level of attention to each word vector. In Figure 7 we present the reference and predicted summaries, as well as the predicted attention over the first 2 lines for another method in the test set. We do not show the raw code due to space limitations. We observe that our approach can predict the first word "write" correctly, unlike the baseline. We observe that the word "write" gets a higher degree of attention than some surrounding words. We know from the literature that the first predicted action word is very important as encoder-decoder models depend on previous words to generate the next in the sequence [24]. The result is a much better summary overall.

## 7 CONCLUSIONS AND FUTURE WORK

This paper advances the state-of-the-art in two key ways. First, we present a model that learns to predict human attention to source code. We represent "human attention" as the percent total gaze time (*ptgt*, Section 3) that a person reads each token in source code with their eyes. Then, we represent the source code as an abstract syntax tree and design a graph neural network-based model (eye-gnn) to learn to predict *ptgt* for arbitrary source code. In short, we found that eye-gnn is able to learn to predict *ptgt* in a manner consistent with moderate positive correlation with actual, human eye gaze measured by *ptgt*.

Second, we present an augmentation to a "vanilla" neural source code summarization technique that makes use of predicted levels of human attention to code. We create a novel attention mechanism based on human attention in addition to the machine "vanilla" attention. Our mechanism uses eye-gnn to predict *ptgt* for every node in the AST. Then, we combine those predicted *ptgt* values with embedding vectors for the nodes in the AST, to emphasize or de-emphasize those nodes based on the *ptgt* values. We then combine this human attention mechanism with the typical machine attention mechanism. We observe a small but significant improvement over the baseline, which is in line with expectations for biologically-inspired neural models from other domains.

We reiterate that we view this paper as a proof-of-concept in predicting human attention and demonstrating that predicted attention has utility in improving machine attention. We demonstrate this utility in a highly-controlled environment in which we use one typical encoder-decoder model and one augmented version of that model. Although our attention model complements the neural attention layer of this approach, considerable future work is needed to study the effects of augmenting other more complex neural approaches. Demographic difference between participants could also affect the model of mimicked human attention and serves as a possible future direction for research. Note, we only use gaze time to predict human attention, future work is needed to explore temporal coherence such as scanpaths.

To encourage reproducibility and verifiability, we release our dataset, code, models, and supporting artifacts at: https://osf.io/b9sjz [8]

## 8 ACKNOWLEDGEMENT

This work is supported in part by NSF CCF-2100035 and CCF-2211428. Any opinions, findings, and conclusions expressed herein are the authors and do not necessarily reflect those of the sponsors.

## REFERENCES

[1] Nahla J Abid, Jonathan I Maletic, and Bonita Sharif. 2019. Using developer eye movements to externalize the mental model used in code summarization tasks. In *Proceedings of the 11th ACM Symposium on Eye Tracking Research & Applications*. 1–9.

[2] Nahla J Abid, Bonita Sharif, Natalia Dragan, Hend Alrasheed, and Jonathan I Maletic. 2019. Developer reading behavior while summarizing java methods: Size and context matters. In *2019 IEEE/ACM 41st International Conference on Software Engineering (ICSE)*. IEEE, 384–395.

[3] Miltiadis Allamanis. 2019. The adverse effects of code duplication in machine learning models of code. In *Proceedings of the 2019 ACM SIGPLAN International Symposium on New Ideas, New Paradigms, and Reflections on Programming and Software*. 143–153.

[4] Miltiadis Allamanis, Marc Brockschmidt, and Mahmoud Khademi. 2018. Learning to represent programs with graphs. *International Conference on Learning Representations* (2018).

[5] Ameer Armaly, Paige Rodeghero, and Collin McMillan. 2017. A comparison of program comprehension strategies by blind and sighted programmers. *IEEE Transactions on Software Engineering* 44, 8 (2017), 712–724.

[6] Ameer Armaly, Paige Rodeghero, and Collin McMillan. 2018. AudioHighlight: Code skimming for blind programmers. In *2018 IEEE International Conference on Software Maintenance and Evolution (ICSME)*. IEEE, 206–216.

[7] Satanjeev Banerjee and Alon Lavie. 2005. METEOR: An automatic metric for MT evaluation with improved correlation with human judgments. In *Proceedings of the acl workshop on intrinsic and extrinsic evaluation measures for machine translation and/or summarization*. 65–72.

[8] Aakash Bansal. 2023. HumanAttn-Artifacts. https://doi.org/10.17605/OSF.IO/B9SJZ

[9] Aakash Bansal, Sakib Haque, and Collin McMillan. 2021. Project-Level Encoding for Neural Source Code Summarization of Subroutines. *International Conference on Program Comprehension* (2021).

[10] Nathaniel Blanchard, Jeffery Kinnison, Brandon RichardWebster, Pouya Bashivan, and Walter J Scheirer. 2018. A neurobiological cross-domain evaluation metric for predictive coding networks. *arXiv preprint arXiv:1805.10726* (2018).

[11] Josh Bongard. 2009. Biologically Inspired Computing. *IEEE computer* 42, 4 (2009), 95–98.

[12] Matthew Botvinick, David GT Barrett, Peter Battaglia, Nando de Freitas, Darshan Kumaran, Joel Z Leibo, Timothy Lillicrap, Joseph Modayil, Shakir Mohamed, and Neil C Rabinowitz. 2017. Building machines that learn and think for themselves. *Behavioral and Brain Sciences* 40 (2017).

[13] Teresa Busjahn, Roman Bednarik, Andrew Begel, Martha Crosby, James H Paterson, Carsten Schulte, Bonita Sharif, and Sascha Tamm. 2015. Eye movements in code reading: Relaxing the linear order. In *2015 IEEE 23rd International Conference on Program Comprehension*. IEEE, 255–265.

[14] Michael L Collard, Michael J Decker, and Jonathan I Maletic. 2011. Lightweight transformation and fact extraction with the srcML toolkit. In *Source Code Analysis and Manipulation (SCAM), 2011 11th IEEE International Working Conference on*. IEEE, 173–184.

[15] Martha E Crosby and Jan Stelovsky. 1990. How do we read algorithms? A case study. *Computer* 23, 1 (1990), 25–35.

[16] Ernest Davis and Gary Marcus. 2017. Causal generative models are just a start. *Behavioral and Brain Sciences* 40 (2017).

[17] Sarah Fakhoury, Devjeet Roy, Harry Pines, Tyler Cleveland, Cole S Peterson, Venera Arnaoudova, Bonita Sharif, and Jonathan I Maletic. 2021. gazel: Supporting Source Code Edits in Eye-Tracking Studies. In *2021 IEEE/ACM 43rd International Conference on Software Engineering: Companion Proceedings (ICSE-Companion)*. IEEE, 69–72.

[18] Andrew Finch, Young-Sook Hwang, and Eiichiro Sumita. 2005. Using machine translation evaluation techniques to determine sentence-level semantic equivalence. In *Proceedings of the third international workshop on paraphrasing (IWP2005)*.

[19] Dario Floreano and Claudio Mattiussi. 2008. *Bio-inspired artificial intelligence: theories, methods, and technologies*. MIT press.

[20] Ruth C Fong, Walter J Scheirer, and David D Cox. 2018. Using human brain activity to guide machine learning. *Scientific reports* 8, 1 (2018), 1–10.

[21] Andrew Forward and Timothy C Lethbridge. 2002. The relevance of software documentation, tools and technologies: a survey. In *Proceedings of the 2002 ACM symposium on Document engineering*. ACM, 26–33.

[22] Xavier Gastaldi. 2017. Shake-shake regularization. *arXiv preprint arXiv:1705.07485* (2017).

[23] Sonia Haiduc, Jairo Aponte, Laura Moreno, and Andrian Marcus. 2010. On the use of automated text summarization techniques for summarizing source code. In *2010 17th Working Conference on Reverse Engineering*. IEEE, 35–44.

[24] Sakib Haque, Aakash Bansal, Lingfei Wu, and Collin McMillan. 2021. Action Word Prediction for Neural Source Code Summarization. *28th IEEE International Conference on Software Analysis, Evolution and Reengineering* (2021).

[25] Sakib Haque, Alexander LeClair, Lingfei Wu, and Collin McMillan. 2020. Improved Automatic Summarization of Subroutines via Attention to File Context. *International Conference on Mining Software Repositories* (2020).

[26] Saihui Hou and Zilei Wang. 2019. Weighted channel dropout for regularization of deep convolutional neural network. In *Proceedings of the AAAI Conference on Artificial Intelligence*, Vol. 33. 8425–8432.

[27] Zhengping Ji and Juyang Weng. 2010. WWN-2: A biologically inspired neural network for concurrent visual attention and recognition. In *The 2010 International Joint Conference on Neural Networks (IJCNN)*. IEEE, 1–8.

[28] Shaohua Jia, Amanda Seccia, Pasha Antonenko, Richard Lamb, Andreas Keil, Matthew Schneps, Marc Pomplun, et al. 2018. Biometric recognition through eye movements using a recurrent neural network. In *2018 IEEE International*

*Conference on Big Knowledge (ICBK)*. IEEE, 57–64.
[29] Brenden M Lake, Tomer D Ullman, Joshua B Tenenbaum, and Samuel J Gershman. 2017. Building machines that learn and think like people. *Behavioral and brain sciences* 40 (2017).
[30] Alexander LeClair, Sakib Haque, Lingfei Wu, and Collin McMillan. 2020. Improved Code Summarization via a Graph Neural Network. In *28th ACM/IEEE International Conference on Program Comprehension (ICPC'20)*.
[31] Alexander LeClair, Siyuan Jiang, and Collin McMillan. 2019. A neural model for generating natural language summaries of program subroutines. In *Proceedings of the 41st International Conference on Software Engineering*. IEEE Press, 795–806.
[32] Alexander LeClair and Collin McMillan. 2019. Recommendations for Datasets for Source Code Summarization. In *Proceedings of the 2019 Conference of the North American Chapter of the Association for Computational Linguistics: Human Language Technologies, Volume 1 (Long and Short Papers)*. 3931–3937.
[33] Shangqing Liu, Yu Chen, Xiaofei Xie, Jing Kai Siow, and Yang Liu. 2021. Retrieval-Augmented Generation for Code Summarization via Hybrid {GNN}. In *International Conference on Learning Representations*. https://openreview.net/forum?id=zv-typ1gPxA
[34] Cristina Melício, Rui Figueiredo, Ana Filipa Almeida, Alexandre Bernardino, and José Santos-Victor. 2018. Object detection and localization with Artificial Foveal Visual Attention. In *2018 Joint IEEE 8th International Conference on Development and Learning and Epigenetic Robotics (ICDL-EpiRob)*. IEEE, 101–106.
[35] Graham Neubig. [n. d.]. Survey of Methods to Generate Natural Language from Source Code. ([n. d.]).
[36] Afonso Nunes, Rui Figueiredo, and Plinio Moreno. 2020. Learning to Search for Objects in Images from Human Gaze Sequences. In *International Conference on Image Analysis and Recognition*. Springer, 280–292.
[37] Unaizah Obaidellah, Mohammed Al Haek, and Peter C.-H. Cheng. 2018. A Survey on the Usage of Eye-Tracking in Computer Programming. *ACM Comput. Surv.* 51, 1 (2018), 5:1–5:58. https://doi.org/10.1145/3145904
[38] André Ofner and Sebastian Stober. 2018. Towards bridging human and artificial cognition: Hybrid variational predictive coding of the physical world, the body and the brain. *Advances in Neural Information Processing Systems* (2018).
[39] Ariyo Oluwasammi, Muhammad Umar Aftab, Zhiguang Qin, Son Tung Ngo, Thang Van Doan, Son Ba Nguyen, Son Hoang Nguyen, and Giang Hoang Nguyen. 2021. Features to Text: A Comprehensive Survey of Deep Learning on Semantic Segmentation and Image Captioning. *Complexity* 2021 (2021).
[40] Simone Palazzo, Francesco Rundo, Sebastiano Battiato, Daniela Giordano, and Concetto Spampinato. 2020. Visual saliency detection guided by neural signals. In *2020 15th IEEE International Conference on Automatic Face and Gesture Recognition (FG 2020)*. IEEE, 525–531.
[41] Simone Palazzo, Concetto Spampinato, Isaak Kavasidis, Daniela Giordano, and Mubarak Shah. 2018. Decoding Brain Representations by Multimodal Learning of Neural Activity and Visual Features. *arXiv preprint arXiv:1810.10974* (2018).
[42] Matteo Paltenghi and Michael Pradel. 2021. Thinking like a developer? comparing the attention of humans with neural models of code. In *2021 36th IEEE/ACM International Conference on Automated Software Engineering (ASE)*. IEEE, 867–879.
[43] Kishore Papineni, Salim Roukos, Todd Ward, and Wei-Jing Zhu. 2002. BLEU: A Method for Automatic Evaluation of Machine Translation. In *Proceedings of the 40th Annual Meeting on Association for Computational Linguistics* (Philadelphia, Pennsylvania) *(ACL '02)*. Association for Computational Linguistics, Stroudsburg, PA, USA, 311–318. https://doi.org/10.3115/1073083.1073135
[44] Vasili Ramanishka, Abir Das, Jianming Zhang, and Kate Saenko. 2017. Top-down visual saliency guided by captions. In *Proceedings of the IEEE conference on computer vision and pattern recognition*. 7206–7215.
[45] Keith Rayner. 2009. Eye movements and attention in reading, scene perception, and visual search. *The quarterly journal of experimental psychology* 62, 8 (2009), 1457–1506.
[46] Paige Rodeghero, Cheng Liu, Paul W McBurney, and Collin McMillan. 2015. An eye-tracking study of java programmers and application to source code summarization. *IEEE Transactions on Software Engineering* 41, 11 (2015), 1038–1054.
[47] Paige Rodeghero and Collin McMillan. 2015. An empirical study on the patterns of eye movement during summarization tasks. In *2015 ACM/IEEE International Symposium on Empirical Software Engineering and Measurement (ESEM)*. 1–10.
[48] Paige Rodeghero, Collin McMillan, Paul W. McBurney, Nigel Bosch, and Sidney D'Mello. 2014. Artifacts for Improving Automated Source Code Summarization via an Eye-Tracking Study of Programmers. https://notredame.box.com/s/bhyoqle1i90vuz75pu4ufch3rpz1lpij
[49] Paige Rodeghero, Collin McMillan, Paul W McBurney, Nigel Bosch, and Sidney D'Mello. 2014. Improving automated source code summarization via an eye-tracking study of programmers. In *Proceedings of the 36th international conference on Software engineering*. ACM, 390–401.
[50] Devjeet Roy, Sarah Fakhoury, and Venera Arnaoudova. 2021. Reassessing Automatic Evaluation Metrics for Code Summarization Tasks. In *Proceedings of the ACM Joint European Software Engineering Conference and Symposium on the Foundations of Software Engineering (ESEC/FSE)*.

[51] Aakash Bansal Sakib Haque, Zachary Eberhart and Collin McMillan. 2022. Semantic Similarity Metrics for Evaluating Source Code Summarization. In *30th International Conference on Program Comprehension (ICPC'22)*.
[52] Zohreh Sharafi, Bonita Sharif, Yann-Gaël Guéhéneuc, Andrew Begel, Roman Bednarik, and Martha E. Crosby. 2020. A practical guide on conducting eye tracking studies in software engineering. *Empir. Softw. Eng.* 25, 5 (2020), 3128–3174. https://doi.org/10.1007/s10664-020-09829-4
[53] Zohreh Sharafi, Zéphyrin Soh, and Yann-Gaël Guéhéneuc. 2015. A systematic literature review on the usage of eye-tracking in software engineering. *Information and Software Technology* 67 (2015), 79–107.
[54] Bonita Sharif, Michael Falcone, and Jonathan I Maletic. 2012. An eye-tracking study on the role of scan time in finding source code defects. In *Proceedings of the Symposium on Eye Tracking Research and Applications*. 381–384.
[55] Bonita Sharif and Huzefa Kagdi. 2011. On the use of eye tracking in software traceability. In *Proceedings of the 6th International Workshop on Traceability in Emerging Forms of Software Engineering*. 67–70.
[56] Bonita Sharif, John Meinken, Timothy Shaffer, and Huzefa H. Kagdi. 2017. Eye movements in software traceability link recovery. *Empir. Softw. Eng.* 22, 3 (2017), 1063–1102. https://doi.org/10.1007/s10664-016-9486-9
[57] Fabian H Sinz, Xaq Pitkow, Jacob Reimer, Matthias Bethge, and Andreas S Tolias. 2019. Engineering a Less Artificial Intelligence. *Neuron* 103, 6 (2019), 967–979.
[58] Jamie Starke, Chris Luce, and Jonathan Sillito. 2009. Searching and skimming: An exploratory study. In *2009 IEEE International Conference on Software Maintenance*. IEEE, 157–166.
[59] James Tanner and Laurent Itti. 2019. A top-down saliency model with goal relevance. *Journal of vision* 19, 1 (2019), 11–11.
[60] Li Wan, Matthew Zeiler, Sixin Zhang, Yann Le Cun, and Rob Fergus. 2013. Regularization of neural networks using dropconnect. In *International conference on machine learning*. PMLR, 1058–1066.
[61] Fengrong Zhao, Junqi Zhao, and Yang Bai. 2020. A Survey of Automatic Generation of Code Comments. In *Proceedings of the 2020 4th International Conference on Management Engineering, Software Engineering and Service Sciences*. 21–25.
[62] Daniel Zügner, Tobias Kirschstein, Michele Catasta, Jure Leskovec, and Stephan Günnemann. 2021. Language-Agnostic Representation Learning of Source Code from Structure and Context. In *International Conference on Learning Representations*.

## VERSION HISTORY